\documentclass[seceq]{ptptex}

\usepackage{graphicx}
\notypesetlogo               

\newcommand{\nc}{\newcommand}		
\newcommand{\renc}{\renewcommand}	
\nc{\nuc}[2]	{$^{#1}${#2}} 		
\nc{\vc}[1]	{\mbox{\boldmath $#1$}}	
\nc{\bra}	{\langle}		
\nc{\ket}	{\rangle}		
\nc{\wtil}	{\widetilde}		
\nc{\del}       {\partial}
\nc{\mapleft}[1]{			
 \smash{\mathop{\,			%
  \hbox to 1.2cm{\rightarrowfill}\, }\limits_{#1}}}

\begin{document}

\markboth{T. Myo, S. Sugimoto, K. Kat\=o, H. Toki and K. Ikeda}{   
Tensor Correlation in $^4$He with the Tensor-Optimized Shell Model
}

\title{
Tensor Correlation in $^4$He with the Tensor-Optimized Shell Model
}

\author{
Takayuki \textsc{Myo},$^{1,}$\footnote{\noindent E-mail : myo@rcnp.osaka-u.ac.jp}
Satoru \textsc{Sugimoto},$^{2,}$\footnote{\noindent E-mail : satoru@ruby.scphys.kyoto-u.ac.jp}
Kiyoshi \textsc{Kat\=o}$^{3,}$\footnote{\noindent E-mail : kato@nucl.sci.hokudai.ac.jp}
Hiroshi \textsc{Toki},$^{1,}$\footnote{\noindent E-mail : toki@rcnp.osaka-u.ac.jp}
and
Kiyomi \textsc{Ikeda}$^4,$\footnote{\noindent E-mail : k-ikeda@postman.riken.go.jp}
}

\inst{
$^1$Research Center for Nuclear Physics (RCNP), Osaka University,\\ Ibaraki 567-0047, Japan\\
$^2$Department of Physics, Graduate School of Science, Kyoto University,\\ Kyoto 606-8502, Japan\\
$^3$Division of Physics, Graduate School of Science, Hokkaido University,\\ Sapporo 060-0810, Japan\\
$^4$The Institute of Physical and Chemical Research (RIKEN),\\ Wako 351-0198, Japan.
}

\recdate{May 2, 2006}

\abst{
We study the characteristics of the tensor correlation in $^4$He using a shell model type method.  We treat the tensor force explicitly by performing a configuration-mixing calculation in the $2p2h$ basis and include single-particle states up to intermediately high angular momenta. We adopt the Gaussian expansion method for the quantitative description 
of the spatial shrinkage of the single-particle states to optimize the tensor correlation.
We are able to describe the full strength of the tensor correlation for $^4$He in the shell model type method by realizing convergence. We call this model the tensor-optimized shell model. 
It is found that in $^4$He, three specific $2p2h$ configurations are strongly coupled 
with the $(0s)^4$ configuration due to the characteristic features of the tensor operator.
}

\maketitle

\section{Introduction}

The tensor force is an essential component of the nuclear force
and plays a significant role in nuclear structure. 
For $^4$He, accurate calculations using a realistic nucleon-nucleon interaction demonstrated that the contribution of the tensor force is very large, reaching $-68$ MeV, 
which is of the same order as or even larger than that of the central force\cite{Ak86,No00,Ka01}.
The resulting strong tensor correlation enhances the $D$-state probability up to about 15\% in the wave function.  In light nuclei up to around mass number 10 ($A \leq 10$), 
the Green's function Monte Calro (GFMC) method\cite{Wi00} was developed to treat realistic interactions and is capable of describing ground states and a few excited states. With this method, it was demonstrated that the contribution of the one-pion exchange potential (OPEP) provides 70\%-80\% of the two-body attraction, in which the dominant component of OPEP is the tensor force.

We wish to describe nuclei with larger mass numbers explicitly using realistic nucleon-nucleon interactions.
Variational methods including GFMC of few-body systems use the relative coordinates of nucleons, which are suitable to work out the nucleon-nucleon interaction.  We call this the many-body theory with $T$-type basis, because the two interacting nucleons, whose centers of mass are connected by some reference coordinate, interact directly through the nucleon-nucleon interaction.  In this case, the number of relative coordinates increase as $A(A-1)/2$, where $A$ is the number of nucleons.  Hence, the $T$-type basis is not suited for heavier systems, and calculations become increasingly difficult as $A$ increases.
By contrast, the mean-field framework uses as the coordinates the positions of the nucleons relative to the center of the nucleus. In this case, the number of coordinates is $A$, and hence this framework is suited for heavy systems.  We call this method the many-body theory with $V$-type basis, since the two interacting nucleons are labeled by the coordinates representing the distances from the center of the nucleus, and they interact only indirectly through the nucleon-nucleon interaction.  Hence, it is difficult to treat realistic interactions involving short-range repulsion and the tensor force, because many configurations in the $V$-type basis are needed to describe motion under such a realistic interaction. In this case, we have to invent some method to treat these features of realistic interactions.

The standard method for the description of many-body systems in the $V$-type basis is the Brueckner-Hartree-Fock (BHF) method,\cite{Mu00,Ak72,An79} in which two-body interactions with a short-range repulsion and a strong tensor force are treated with the Brueckner $G$-matrix effective interaction under the independent pair approximation.  Since the resulting $G$-matrix effective interaction is a smooth interaction, the wave functions in the $V$-type basis can treat the $G$-matrix within the model space.  The BHF method or the shell model, which is based on the Brueckner theory, have some success in the description of many-body systems.  However, there seem to be some essential features missing, such as the spin-orbit interaction\cite{An81,Na03} and we are therefore forced to employ some phenomenology.

Recently, there have been two important steps proposed for the full description of nuclei in the $V$-type basis.  One is the method of Neff and Feldmeier which treats the unitary transformation of the radial correlation due to the short-range repulsion and the tensor correlation due to the tensor force separately\cite{Ne03}.  The unitary transformation method introduces the unitary operator for the short-range repulsion and another unitary operator for the tensor force.  It seems that the short-range repulsion is treated properly, because it acts at a very small distance ($r\le 0.5$ fm), and therefore the terms more than the three-body operators caused by the radial unitary transformations can be ignored\cite{Fe98}.  By contrast, the tensor force is not treated properly within the two-body terms of the unitary transformed Hamiltonian.  This may be related to the fact that the range of the tensor force is intermediate (0.5 fm $\le r \le$1.4 fm). 

The second proposed method is that of Toki et al.\cite{To02} for the treatment of the pion exchange interaction in terms of the relativistic mean-field framework.  The mean-field method was introduced in the relativistic mean-field approximation with a finite pion mean field.  Because the pion is a pseudo-scalar meson, the finite pion mean field in the spherical ansatz naturally induces parity mixing of the intrinsic single-particle states.  This relativistic mean-field model provides an interesting mechanism for the splitting of the spin-orbit partner due to the mixing of the parity partners with the same total spin as $s_{1/2}$ and $p_{1/2}$ or $p_{3/2}$ and $d_{3/2}$\cite{Og04}.  The results of this work led us to consider employing the parity mixing single-particle states for the description of the tensor correlation in the non-relativistic approximation.  Furthermore, it is straightforward to perform the parity projection and the charge projection, due to the fact that the tensor force is isospin dependent\cite{Su04}.  The leading term in the parity and charge projection consists of $2p2h$ configurations\cite{To02}.  

With the results described above, the next step was to employ a shell model type prescription in which the closed shell wave function is treated as the $0p0h$ state and $2p2h$ wave functions are introduced to treat the tensor force\cite{My05}. When this method is applied to $^4$He, for which the closed shell state is $(0s_{1/2})^4$ and the particle states of the $2p2h$ configurations are those in the $p$-orbit, the major $2p2h$ state is $(0p_{1/2})^2(0s_{1/2})^{-2}$, pionic state with the $0^-$ coupling in the particle-hole picture,\cite{To02} and the spatial extension of the $0p_{1/2}$ wave function is almost half that of the ordinary harmonic oscillator wave function\cite{My05}.  In this case, however, we still had to strengthen the matrix elements of the tensor force by 50\% in order to provide a sufficiently large amount of the $D$-state probability of $^4$He.  We have to understand why we ought to increase the tensor matrix elements by 50\%.

In this paper, we would like to extend further the shell model type method by increasing the number of $2p2h$ configurations in order to understand the origin of the 50\% increase of the tensor matrix element.  Let us consider more carefully the characteristics of the tensor force.  The tensor force is $V_T=f_T(r) S_{12}$, with the tensor operator $S_{12}=\sqrt{24\pi} [Y_2(\hat r_{12}),~[\vc{\sigma}^1,\vc{\sigma}^2]_2 ]_0$.  Both the coupled spin ($S$) and the coupled orbital angular momentum ($L$) of two particles must flip with $\Delta S=\Delta L=2$.  Hence, the lowest relative angular momentum state for two particles is the $L=2$ state, for which  the centrifugal potential is given by $V_{\rm cen}= 6\hbar^2/(m_N r^2)$, and it is 1 GeV at 0.5 fm.  This means we should describe the distance as short as about 0.5 fm and larger by introducing higher angular momentum states.  We have the relation $f_T(r)=\sum_l f_l(r_1,r_2)P_l(\hat r_{12})$.  Here, $P_l({\hat r_{12}})$ has a significant strength around $\hat r \sim 0$,  with a width of about $1/l$ for large $l~(l \ge 5$).  We can write $P_l(\hat r_{12})=(4\pi /\sqrt{2l+1})(-)^l[Y_l(\hat r_1), Y_l(\hat r_2)]_0$.  Because the relative distance to be expressed in terms of the high-spin state with $l$ of the $V$-type coordinate is $r_1 \times (1/l) \ge$ 0.5 fm, with $r_1\sim r_2 \sim 2$ fm (the size of $^4$He), we find $l \le 4$.  In addition, there is an additional change of the angular momentum by 2 for the case of the tensor force.  This means that we may be able to describe the tensor matrix elements by taking the single-particle states with angular momentum up to $l=4+2=6$ for the $^4$He case.  Hence, we may be able to describe the tensor force in terms of the $V$-type coordinates with intermediate size angular momenta.  It remains to construct a shell model type prescription to handle the tensor force by using the $V$-type basis.  We name this model the tensor optimized shell model.

The tensor optimized shell model (TOSM) is described as follows.\\
(1) We base it on a closed shell state, like the $0p0h$ state, and add $2p2h$ states.\\
(2) We include intermediate high angular momentum orbits for the particle states within the $2p2h$ excitations.\\
(3) We allow radial shrinkage of the particle states by employing the Gaussian wave function with the lowest node for each angular momentum state.\\
(4) We use the many-body Hamiltonian with a suitably chosen two-body interaction that explicitly includes the tensor force.\\
For $^4$He, we take the $(0s_{1/2})^4$ state as the basic state and add $2p2h$ states with the particle states of finite angular momenta as large as the result converges, with convergence expected to occur around $l=6$.  No angular momentum states have nodes, except the $s$-state, which is constructed as a state orthogonal to the $0s$ state with Gaussian wave functions with various size parameters. We do not consider the problem for the case of a short-range repulsion in this paper and instead use one obtained phenomenologically.  
However, once we have succeeded in describing the tensor force, we plan to also determine the short-range part.

In \S\ref{sec:model}, we describe in detail the tensor-optimized shell model (TOSM), which treats the tensor correlation explicitly. 
In \S\ref{sec:result} we investigate the structure of $^4$He by considering high angular momentum states and discuss the characteristics of the tensor correlation.
In \S\ref{sec:gauss}, we study the tensor correlation in detail in terms of the radial shrinkage with the Gaussian expansion method.
In \S\ref{sec:tens}, we treat $^4$He by employing the bare tensor force of AV8$^\prime$ with a slight modification of the central interaction in the TOSM.
The present study is summarized in \S\ref{sec:summary}.

\section{Tensor optimized shell model for $^4$He}\label{sec:model}

In this paper, we apply the tensor optimized shell model to $^4$He.  Particularly, we would like to see if the tensor correlation is satisfactorily described by employing Gaussian wave functions with intermediate size angular momenta.

\subsection{Higher partial waves and variational methods}

We base $2p2h$ wave functions in addition to the closed shell wave function as the $0p0h$ state and adopt the $(0s_{1/2})^4$ state for $^4$He.  We include $2p2h$ excitations with higher partial waves up to $l_{\rm max}$, the maximum orbital angular momentum of the excited particle orbits.
We successively increase the value of $l_{\rm max}$ and watch the convergence of the solutions.
The $1s$ orbit is included in the cases $l_{\rm max}\geqq2$.
We describe the particle states while maintaining their orthogonality to the occupied states.
We do not include higher nodal orbits, such as $2s$ and $1p$, except for the $1s$ orbit.
The reason is as follows; We have confirmed that with the usual shell model type calculation 
including the higher nodal orbits of harmonic oscillator wave functions (HOWF) with a common length parameter, 
it is difficult to describe the tensor correlation satisfactorily.  
In fact, the tensor correlation was investigated using a higher nodal HOWF in a previous study,\cite{Sh73} and it was shown that the convergence of the solutions with respect to the principal quantum number $N$ is slow.
Furthermore, we found that the tensor correlation is optimized with the small size parameters of the Gaussian wave functions for the excited particle states.
These results imply that the ordinary HOWF is not sufficient to satisfactorily represent the tensor correlation.
For this reason, instead of HOWF, we adopt the Gaussian expansion technique\cite{Hi03,Ao95} for single-particle orbits.
Each Gaussian basis function has the form of a nodeless HOWF, except for $1s$ orbit, and 
when we superpose a sufficient number of Gaussian basis functions with appropriate length parameters, we can fully optimize the radial component of every orbit.

The form of the Gaussian basis function for the label $\alpha$, such as $0s_{1/2}$ and $0p_{1/2}$, is expressed as
\begin{eqnarray}
	\phi_{\alpha}(\vc{r},b_{\alpha,m})
&=&	N_l(b_{\alpha,m})\ r^l\ e^{-(r/b_{\alpha,m})^2/2}\ [Y_{l}(\hat{\vc{r}}),\chi_{1/2}]_j,
	\label{eq:Gauss2}
	\\
	N_l(b_{\alpha,m})
&=&	\left[	\frac{2\ b_{\alpha,m}^{-(2l+3)} }{ \Gamma(l+3/2)}\right]^{\frac12},
\end{eqnarray}
where $m$ is an index that distinguish Gaussian basis functions with different values of the length parameter $b_{\alpha,m}$.
We construct the following ortho-normalized single-particle wave function $\psi^n_{\alpha}$
with a linear combination of Gaussian bases 
\begin{eqnarray}
	\psi^n_{\alpha}(\vc{r})
&=&	\sum_{m=1}^{N_\alpha} d^n_{\alpha,m}\ \phi_{\alpha}(\vc{r},b_{\alpha,m})
	\qquad
	{\rm for}~~n~=~1,\cdots,N_\alpha.	
	\label{eq:Gauss1}
\end{eqnarray}
Here, $N_\alpha$ is a number of basis functions for $\alpha$.
The coefficients $\{d^n_{\alpha,m}\}$ are determined by solving the 
eigenvalue problem for the norm matrix of the non-orthogonal Gaussian basis set given in Eq.~(\ref{eq:Gauss2}).
We obtain the new single-particle wave functions $\{\psi^n_{\alpha}\}$ in Eq.~(\ref{eq:Gauss1})
which are distinguished by the index $n$ for different values of the radial component with a given label $\alpha$. 

We choose the Gaussian basis functions for the particle states to be orthogonal to the occupied single-particle states, which are $0s_{1/2}$ in the case of $^4$He. For $0s_{1/2}$ states, we simply employ one Gaussian basis function, namely, HOWF with length $b_{0s_{1/2},m=1}=b_{0s}$. 
For $1s_{1/2}$ states, we introduce a basis orthogonal to the $0s_{1/2}$ states
and possessing a length parameter $b_{1s,m}$ that differs from $b_{0s}$. This extended $1s_{1/2}$ state is expressed as
\begin{eqnarray}
	\phi_{1s}(\vc{r},b_{0s},b_{1s,m})
&=&	N(b_{1s,m})\ \{ f_0(b_{0s},b_{1s,m}) + f_2(b_{0s},b_{1s,m})\ r^2 \}\ 
	\nonumber
	\\
&\times&e^{-(r/b_{1s,m})^2/2} \ Y_{00}(\hat{\vc{r}}),
	\label{eq:1s}
\end{eqnarray}
where we omit the spin wave function. Condition of coefficients $f_0$ and $f_2$ are determined by 
the normalization of $\phi_{1s}$ and the orthogonality to the $0s$ state.  The ortho-normalized $1s$ wave functions $\{\psi^n_{1s}\}$ are obtained from $\phi_{1s}$ with various range parameters, as described above.

Here we set the variational wave function for $^4$He.
Each configuration of $^4$He is expressed as anti-symmetrized products of the four single-particle states,
which are distinguished by $\alpha$ and the basis index $n$, as
\begin{eqnarray}
    \Phi_{p}
~=~ {\cal A}\bigg\{\prod_{k=1}^{4} \psi^{n_k}_{\alpha_k} \bigg\},
\end{eqnarray}
where $p$ indicates the set $\{\alpha_k, n_k; k=1,\cdots,4\}$. 
The total wave function of $^4$He with $(J^\pi,T)=(0^+,0)$ for spin and isospin
is expressed a superposition of wave function $\Phi_{p}$ as
\begin{eqnarray}
    \Psi(^{4}{\rm He})
&=& \sum_{p}\ a_{p}\: \Phi_{p},\
    \label{eq:WF1}
\end{eqnarray}
where $a_{p}$ represents variational coefficients.

The variation of the energy expectation value with respect to 
the total wave function $\Psi(^{4}{\rm He})$ is given by
\begin{eqnarray}
\delta\frac{\bra\Psi|H|\Psi\ket}{\bra\Psi|\Psi\ket}&=&0\ ,
\end{eqnarray}
which leads to the following equations:
\begin{eqnarray}
    \frac{\del \bra\Psi| H - E |\Psi \ket} {\del b_{\alpha,m}}
&=& 0\ ,\quad
    \frac{\del \bra\Psi| H - E |\Psi \ket} {\del a_{p}}
=   0\ .
   \label{eq:vari}
\end{eqnarray}
Here, $E$ is a Lagrange multiplier corresponding to the total energy. 
The parameters $\{b_{\alpha,m}\}$ for the Gaussian bases appear in non-linear forms in the energy expectation value.
We solve these two kinds of variational equations in the following steps. 
First, fixing all the length parameters $b_{\alpha,m}$, we solve the linear equation for $\{a_{p}\}$ 
as an eigenvalue problem for $H$ with partial waves up to $l_{\rm max}$. 
We thereby obtain the eigenvalue $E$, which is a function of $\{b_{\alpha,m}\}$. 
Next, we try various sets of the length parameters $\{b_{\alpha,m}\}$ 
in order to find the solution which minimizes the total energy. 
In this wave function, we can describe the spatial shrinkage with an appropriate radial form,
which is important for the tensor correlation.

The analysis carried out in this paper consists of two steps.
The first is the main analysis in which we prepare a single Gaussian basis function for each label $\alpha$ [$N_\alpha=1$ in Eq.~(\ref{eq:Gauss1})].
In this case, we study the convergence of the energy by including higher partial waves 
and investigate the characteristics of the configuration mixing. 
As the second part of the analysis, we adopt the Gaussian expansion method for the excited nucleon states and examine 
the quantitative description of the spatial shrinkage of the wave function.

Our wave function may contain an excitation of the spurious center of mass (cm) motion.
We estimate the amount of this excitation using a following operator $H_G$;
\begin{eqnarray}
	H_G
&=&	T_G~+~\frac12\, M_G\, \omega^2\, \vc{R}_G^2,\qquad
	T_G
\,=\,	\frac{\vc{P}_G^2}{2M_G}
	\label{eq:CM}
	\\
	M_G
&=&	A\cdot m_N,\quad
	\vc{R}_G
\,=\, 	\frac1A\, \sum_{i=1}^A \vc{r}_i,\quad
	\vc{P}_G
\,=\,	\sum_{i=1}^A \vc{p}_i,\quad
	\omega
\,=\,	\frac{\hbar}{m_N\, b_{0s}^2},
\end{eqnarray}
where $A$ is a mass number.
When the c.m. motion is constrained to the $0s$ state in the total wave function,
$\bra H_G \ket$ must be $1.5\, \hbar\omega$.
We estimate the amount of the spurious cm motion 
by considering the deviation of $\bra H_G \ket$ from $1.5 \hbar\omega$.

\subsection{Hamiltonian}

The Hamiltonian for $^4$He is given by
\begin{eqnarray}
    H
&=& \sum_{i=1}^4{t_i} - T_G
+   \sum_{i<j}^4 v_{ij}\ ,
    \label{eq:Ham}
\end{eqnarray}
with
\begin{eqnarray}
    v_{ij}
&=& v_{ij}^C + v_{ij}^{T} + v_{ij}^{LS} + v_{ij}^{Clmb}\ .
\end{eqnarray}
The effective $NN$ interaction $v_{ij}$ consists of central ($v^C_{ij}$), tensor ($v^T_{ij}$), LS
($v^{LS}_{ij}$) and Coulomb ($v_{ij}^{Clmb}$) terms.
We do not explicitly treat the short-range correlation and instead use some phenomenological
central interactions.  For the tensor force, we use one of the effective tensor forces in which the short-range part is renormalized, 
as this has been the standard choice for our study of the tensor correlation.
We demonstrate toward the end of this paper, however, that the tensor correlation does not depend on the effective tensor interaction qualitatively, because of the strong centrifugal potential.\cite{My05}

We next explain effective interactions that include the tensor force explicitly and are suitable for the present analysis.
We use a sufficiently strong tensor force comparable to realistic interactions,
because we would like to determine the characteristics of the tensor correlation in the structure of $^4$He.
In Ref.\citen{My05}, we examined two kinds of effective interaction, AK\cite{Ak04,Ik04} and GPT\cite{Go70}. 
The former is constructed from the AV8$^\prime$ realistic interaction using the $G$-matrix theory,
where the cutoff momentum of the $Q$-space for the $G$-matrices
is chosen as $k_Q=2.8$ fm$^{-1}$, twice the usual Fermi momentum ($k_F=1.4$ fm$^{-1}$). 
In this prescription, short-range correlations including very high momentum components ($k>k_Q$) of
the tensor correlations are renormalized into the central term of the $G$-matrices. 
The tensor force giving momentum components with $k<k_Q$ in the $G$-matrices
survives in intermediate and long ranges\cite{Ak04,Ik04}.
In fact, the matrix element of the tensor force given by AK is very close to that obtained using the bare case (AV8$^\prime$),
which is discussed in \S\ref{sec:tens}.
We also consider the so-called GPT interaction,\cite{Go70} which has a mild short-range repulsion
and was constructed to reproduce the two-nucleon properties.
In GPT, renormalization of the tensor force is regarded as being based on 
the conventional approach, which gives a weak tensor force.

We found that AK gives a large tensor contribution in $^4$He,\cite{My05} 
and it seems to be suitable for the present model to investigate the tensor correlation.
However, as shown below, this force gives a small radius and results in an overbinding problem for $^4$He.
Contrastingly, GPT gives a good radius and a good binding energy, but a small tensor contribution.\cite{My05} 
It is necessary to construct other appropriate interactions 
that take into account the short-range and tensor correlations consistently for the present model.
This is, however, beyond the scope of the present paper.
In this study, we phenomenologically improve AK, with the central part replaced by that of GPT; we call this interaction GA. 
With this interaction, the total energy and the radius are improved in $^4$He\cite{My05}.
Thus, GA can be regarded as a phenomenological effective interaction, 
and we use this interaction to study the tensor correlation for $^4$He.

\section{Results of tensor correlation in the TOSM}\label{sec:result}

We present here the results of the configuration mixing in TOSM and the contributions of higher angular momentum states for the tensor correlation in $^4$He.  
We first present the calculated results with partial waves for $l_{\rm max}$=1 to 6,
with a single Gaussian basis ($N_\alpha=1$) for each single-particle state $\psi_\alpha$,
where we optimize the length parameters of the Gaussian basis $\{b_{\alpha,m=1}\}=\{b_\alpha\}$ for every state.  
We also discuss the characteristic features of the tensor correlation.

\subsection{The calculated results with intermediate high angular momentum states}

Here we present the results for the energy minimum point in each case of $l_{\rm max}$.  
The results for the set $\{b_\alpha\}$ are listed in Table~\ref{tab:1} 
and for the properties of the $^4$He wave functions in Table \ref{tab:2}. 
From Table~\ref{tab:1}, the energy minima are found to satisfy $b_{\alpha\neq 0s}\sim 0.6\ b_{0s}$,
which indicates spatial shrinkage of the wave function.
As discussed below, the expectation value of the tensor force $\langle V_T \rangle$ gives the largest contribution around this energy minimum.

\begin{table}[b]
\begin{center}
\caption{The optimized length parameters of single-particle states for $^4$He in unit of fm.}
\renc{\baselinestretch}{1.20}
\label{tab:1}
\begin{tabular}{l|cccccccc}\hline\hline
$l_{\max}$& $0s_{1/2}$ & $0p_{1/2}$ & $0p_{3/2}$ & $1s_{1/2}$ & $0d_{3/2}$ & $0d_{5/2}$ & $0f_{5/2}$ & $0f_{7/2}$ \\ \hline
1         &  1.26      & 0.75       & 0.69       & ---        & ---        & ---        &   ---      &   ---      \\ 
2         &  1.19      & 0.78       & 0.75       & 0.76       & 0.69       & 0.62       &   ---      &   ---      \\ 
3         &  1.16      & 0.74       & 0.66       & 0.73       & 0.67       & 0.62       &   0.77     &  0.66      \\ 
4         &  1.16      & 0.75       & 0.67       & 0.73       & 0.67       & 0.61       &   0.77     &  0.67      \\ 
5         &  1.16      & 0.76       & 0.67       & 0.73       & 0.67       & 0.61       &   0.77     &  0.64      \\ 
6         &  1.16      & 0.76       & 0.67       & 0.73       & 0.67       & 0.61       &   0.77     &  0.64      \\ \hline
\end{tabular}
\begin{tabular}{l|cccccc}\hline\hline
$l_{\rm max}$ & $0g_{7/2}$ & $0g_{9/2}$ & $0h_{9/2}$ & $0h_{11/2}$ & $0i_{11/2}$ & $0i_{13/2}$ \\ \hline
4         &  0.70      &  0.67      &  ---       &  ---        &  ---       &  ---        \\ 
5         &  0.70      &  0.68      &  0.70      &  0.67       &  ---       &  ---       \\ 
6         &  0.70      &  0.67      &  0.70      &  0.69       &  0.71      &  0.65       \\ \hline
\end{tabular}
\end{center}
\end{table}

\begin{figure}[b]
\centering
\includegraphics[width=9.5cm,clip]{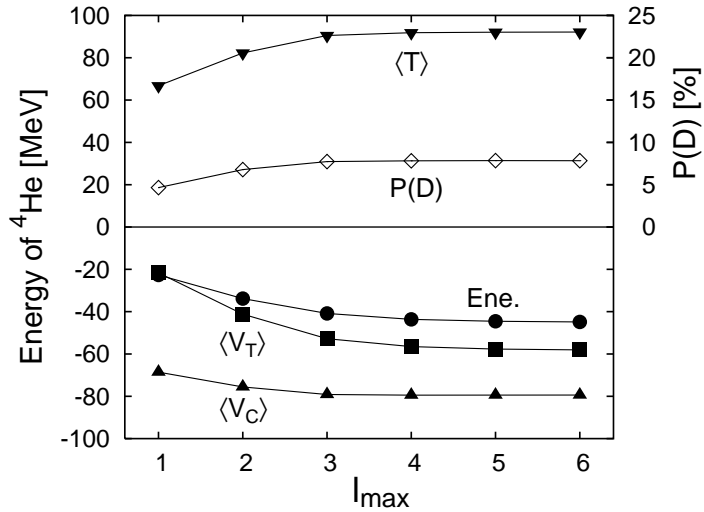}
\caption{Convergence of the properties of \nuc{4}{He} with respect to $l_{\rm max}$.}
\label{fig:ene1}
\end{figure}

From Table~\ref{tab:2} and Fig.~\ref{fig:ene1}, with $l_{\rm max}$ we see the convergence of the energy, 
the central force contribution $\langle V_C \rangle$, $\langle V_T \rangle$, and the $D$-state probability $P(D)$.
Here, we define $P(D)$ as the component for which the values of the total orbital angular momentum and total spin 
are both 2 in the total wave function $\Psi(^{4}{\rm He})$ in Eq.~(\ref{eq:WF1}).
It is also found that the convergence of $\langle V_T \rangle$ is slower than that of $\langle V_C \rangle$.
This implies that the higher partial waves are necessary for the tensor correlation.
For $\langle V_T \rangle$ and $P(D)$, the $l_{\rm max}=4$ case is almost sufficient in the present model.
The excitation of the spurious c.m. motion is small, as judged from the values of $\langle H_G \rangle$, 
and decreases with $l_{\rm max}$ as shown in Table \ref{tab:2}. 
 
\begin{table}[b]
\begin{center}
\caption{The properties of \nuc{4}{He} for each $l_{\rm max}$.
We list the contributions from each term in the Hamiltonian (in units of MeV) and $H_G$ (in 1.5 $\hbar\omega$), the matter radii ($R_m$, in fm)
and $P(D)$ (in \%).}
\renc{\baselinestretch}{1.20}
\label{tab:2}
\begin{tabular}{l|cccccc}\hline\hline
$l_{\rm max}$                    &    ~1      &    ~2      &    ~3      &    ~4      &    ~5      &    ~6      \\ \hline
$E$                              &  $-$22.66  &  $-$33.82  &  $-$40.85  &  $-$43.65  &  $-$44.55  &  $-$44.85  \\
$\bra T\ket$                     & ~~\,66.77  & ~~\,82.25  & ~~\,90.53  & ~~\,91.83  & ~~\,92.08  & ~~\,92.16  \\
$\bra V_{C}\ket$                 &  $-$68.59  &  $-$75.57  &  $-$79.16  &  $-$79.46  &  $-$79.43  &  $-$79.40  \\
$\bra V_{T}\ket$                 &  $-$21.65  &  $-$41.17  &  $-$52.80  &  $-$56.52  &  $-$57.67  &  $-$58.06  \\
$\bra V_{LS}\ket$                & ~\,$-$0.67 & ~\,$-$0.20 & ~\,$-$0.30 & ~\,$-$0.36 & ~\,$-$0.39 & ~\,$-$0.40 \\
$\bra V_{Clmb}\ket$              &~~~\,0.87   &~~~\,0.87   &~~~\,0.88   &~~~\,0.87   &~~~\,0.86   &~~~\,0.86   \\ 
$R_m$                            &~~~\,1.34   &~~~\,1.27   &~~~\,1.24   &~~~\,1.24   &~~~\,1.24   &~~~\,1.24   \\ 
$P(D)$                           &~~~\,4.66   &~~~\,6.78   &~~~\,7.73   &~~~\,7.82   &~~~\,7.85   &~~~\,7.83   \\ 
$\bra H_{G}\ket/(1.5\hbar\omega$)&~~1.067     &~~1.048     &~~1.024     &~~1.021     &~~1.017     &~~1.016     \\ \hline
\end{tabular}
\end{center}
\end{table}

\begin{table}[t]
\begin{center}
\caption{The configurations of $^4$He in order of decreasing probabilities for each case.}
\renc{\baselinestretch}{1.20}
\label{tab:3}
\begin{tabular}{c|c|c|c|c|c}\hline\hline
  \multicolumn{2}{c|}{$l_{\rm max}$=1}  & \multicolumn{2}{c|}{$l_{\rm max}$=2}    & \multicolumn{2}{c}{$l_{\rm max}$=3}      \\ \hline
$0p0h$                     & 91.61  & $0p0h$                      & 89.38 & $0p0h$                       & 88.53 \\
$(0p_{1/2})^2_{10}$        &  6.16  & $(0p_{1/2})^2_{10}$         &  4.31 & $(0p_{1/2})^2_{10}$          &  3.82 \\ 
$(0p_{3/2})^2_{10}$        &  1.25  & $(1s_{1/2})(0d_{3/2})_{10}$ &  3.00 & $(0p_{3/2})(0f_{5/2})_{10}$  &  2.31 \\ 
$(0p_{1/2})^2_{01}$        &  0.55  & $(0d_{3/2})^2_{10}$         &  0.79 & $(1s_{1/2})(0d_{3/2})_{10}$  &  2.03 \\ 
$(0p_{3/2})^2_{01}$        &  0.37  & $(0p_{3/2})^2_{10}$         &  0.75 & $(0p_{3/2})^2_{10}$          &  0.57 \\ 
$(0p_{1/2})(0p_{3/2})_{10}$&  0.07  & $(0d_{5/2})^2_{10}$         &  0.41 & $(0d_{3/2})^2_{10}$          &  0.54 \\
\hline
\end{tabular}

\begin{tabular}{c|c|c|c|c|c}\hline\hline
  \multicolumn{2}{c|}{$l_{\rm max}$=4}  & \multicolumn{2}{c|}{$l_{\rm max}$=5}    & \multicolumn{2}{c}{$l_{\rm max}$=6}      \\ \hline
   $0p0h$                  & 88.47  & $0p0h$                      & 88.45 & $0p0h$                      & 88.48  \\
$(0p_{1/2})^2_{10}$        &  3.62  & $(0p_{1/2})^2_{10}$         &  3.58 & $(0p_{1/2})^2_{10}$         &  3.56  \\ 
$(1s_{1/2})(0d_{3/2})_{10}$&  2.03  & $(1s_{1/2})(0d_{3/2})_{10}$ &  2.03 & $(1s_{1/2})(0d_{3/2})_{10}$ &  2.03  \\ 
$(0p_{3/2})(0f_{5/2})_{10}$&  1.92  & $(0p_{3/2})(0f_{5/2})_{10}$ &  1.90 & $(0p_{3/2})(0f_{5/2})_{10}$ &  1.90  \\ 
$(0d_{5/2})(0g_{7/2})_{10}$&  0.61  & $(0d_{3/2})^2_{10}$         &  0.54 & $(0d_{3/2})^2_{10}$         &  0.54  \\ 
$(0d_{3/2})^2_{10}$        &  0.54  & $(0d_{5/2})(0g_{7/2})_{10}$ &  0.51 & $(0d_{5/2})(0g_{7/2})_{10}$ &  0.52  \\ 
\hline
\end{tabular}
\end{center}
\end{table}

In Table~\ref{tab:3}, we list values for the six configurations in the order of their probabilities for each case of $l_{\rm max}$.
For $2p2h$ states, excited two-particle states are shown.
The subscripts 00, 01, 10 and 11 represent $J$ and $T$, the spin and isospin for the two-nucleon pair, respectively.
It is found that the three kinds of $2p2h$ configurations with
$(0p_{1/2})^2_{10}$, $[(1s_{1/2})(0d_{3/2})]_{10}$ and $[(0p_{3/2})(0f_{5/2})]_{10}$ for the particle part are significantly mixed.
They all have the values ($J$, $T$)=$(1,0)$, which are the same as those for the deuteron,
and thus this two-nucleon coupling can be understood as a deuteron-like correlation\cite{My05}.

In order to see the relation between the above three configurations and the tensor correlation,
we expand $\langle V_T \rangle$ into two terms of matrix elements
between $0p0h$ and $2p2h$ components and between $2p2h$ and $2p2h$ components as follows:
\begin{eqnarray}
	\langle V_T \rangle
&=&	\sum_{p,p'} a_p \ a_{p'} \  \langle \Phi_p|V_T |\Phi_{p'}\rangle
~=~	\{0p0h\mbox{-}2p2h\}+\{2p2h\mbox{-}2p2h\},
	\label{eq:tensor_expand}
	\\
	\{0p0h\mbox{-}2p2h\}
&\equiv&\sum_{p\in 2p2h} a_{0p0h} \ a_p \  \left( \langle \Phi_{0p0h}|V_T |\Phi_{p}\rangle + \langle \Phi_{p}|V_T |\Phi_{0p0h}\rangle \right),
	\label{eq:tensor_expand1}
	\\
	\{2p2h\mbox{-}2p2h\}
&\equiv&\sum_{p,p'\in 2p2h} a_{p} \ a_{p'} \  \langle \Phi_{p}|V_T |\Phi_{p'}\rangle.
	\label{eq:tensor_expand2}
\end{eqnarray}
We list the contributions of these two terms in Table \ref{tab:00}.
It is found that the first term $\{0p0h\mbox{-}2p2h\}$ is dominant, 
and this result is consistent with that obtained in previous studies.\cite{Ak86}
We furthermore list each component in $\{0p0h$-$2p2h\}$ for the $l_{\rm max}=6$ case in Table \ref{tab:01}. 
Three configurations with large probabilities in Table \ref{tab:3}
give large contributions to $\langle V_T \rangle$, about 70\% of the total value as shown in Table \ref{tab:01}.
This means that three configurations are essential for the tensor correlation in $^4$He.
When we increase $l_{\rm max}$ from 3 to 6, these configurations remain strongly mixed. 
This result indicates that the $l_{\rm max}$=3 case is sufficient to describe the characteristics of 
the configuration mixing for the tensor correlation in $^4$He.

\begin{table}[t]
\centering
\caption{Expansion of $\langle V_T \rangle$ into the two terms in Eqs.~(\ref{eq:tensor_expand}) in units of MeV.}
\renc{\baselinestretch}{1.20}
\label{tab:00}
\begin{tabular}{c|cccccc}\hline\hline
$l_{\rm max}$     &  ~1    &   ~2     &  ~3      &  ~4      &  ~5      & ~6  \\
\hline
$0p0h$-$2p2h$ &  $-$21.02  & $-$42.03 & $-$54.74 & $-$59.25 & $-$60.70 & $-$61.22 \\
$2p2h$-$2p2h$ &~\,$-$0.63  &~~~\;0.86 &~~~\;1.94 &~~~\;2.73 &~~~\;3.03 &~~~\;3.16 \\
\hline
\end{tabular}
\end{table}

\begin{table}[t]
\centering
\caption{Contributions of each $0p0h$-$2p2h$ coupling to $\langle V_T \rangle$ in MeV for the $l_{\rm max}$=6 case.}
\label{tab:01}
\renc{\baselinestretch}{1.20}
\begin{tabular}{c|c}\hline\hline
$2p$                        &  $\langle \Phi_{0p0h}|V_T|\Phi_{2p2h}\rangle$ \\
\hline
$(0p_{1/2})^2_{10}$         &  $-$14.47 \\
$(1s_{1/2})(0d_{3/2})_{10}$ &  $-$15.01 \\
$(0p_{3/2})(0f_{5/2})_{10}$ &  $-$12.00 \\
$(0d_{3/2})^2_{10}$         &  $-$3.67  \\
$(0d_{5/2})(0g_{7/2})_{10}$ &  $-$5.17  \\
\hline
\end{tabular}
\end{table}

\begin{figure}[b]
\centering
\includegraphics[width=13.5cm,clip]{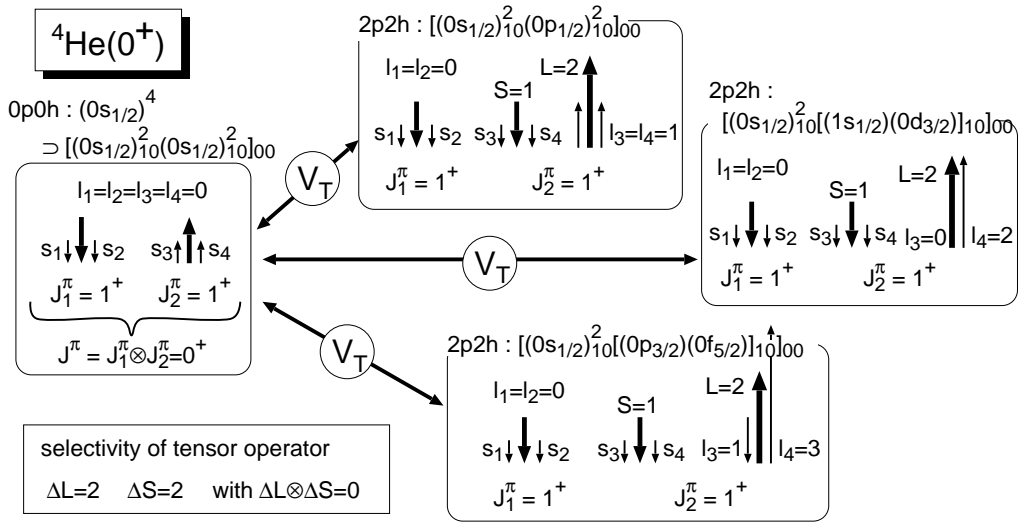}
\caption{Schematic depiction of the favored coupling through the tensor force for $^4$He.}
\label{fig:coupling}
\end{figure}

\subsection{Special feature and the strong shrinkage of the tensor correlation}

We next discuss the reason that three configurations are related to the tensor correlation.
We depict the coupling scheme generated by the tensor force for the $0p0h$ and these $2p2h$ configurations in Fig.~\ref{fig:coupling},
where $\{l_i\}$ and $\{s_i\}$ $(i=1,\cdots,4)$ are the orbital angular momenta and the directions of the intrinsic spins for each nucleon in $^4$He, respectively.
The $0p0h$ configuration of $(0s)^4$ is expressed as
$1/\sqrt{2}\left\{(0s_{1/2})^2_{01}\right.$
\\
$\left.(0s_{1/2})^2_{01}+ (0s_{1/2})^2_{10}(0s_{1/2})^2_{10}\right\}$.
When the tensor force acts on $(0s_{1/2})^2_{10}$ in the $0p0h$ configuration,
it changes both the orbital angular momentum $(L)$ and intrinsic spin $(S)$ for excited particle states by 2 
($\Delta L=2$ and $\Delta S=2$).
The directions of $\Delta L$ and $\Delta S$ must be anti-parallel,
since the rank of the tensor operator $S_{12}\propto[Y_2,[\vc{\sigma}^1,\vc{\sigma}^2]_2]_0$ is zero.
The direction of $S$ in the particle states of the $2p2h$ configuration is also opposite to that of the original $(0s_{1/2})^2_{10}$ component 
in the $0p0h$ case.
If the $L=2$ component in the $2p2h$ configuration consists of $l_3=l_4=1$ for orbital angular momenta of 
two excited nucleons, as shown in Fig.~\ref{fig:coupling},
the directions of the orbital angular momentum and the spin for each excited single-particle state
will be opposite, and hence this state becomes $0p_{1/2}$
and $(0p_{1/2})^2_{10}$ is formed.
The situation discussed here can also be understood as the $0^-$ coupling between $0s_{1/2}$ and $0p_{1/2}$ states
in the particle-hole picture\cite{To02,Su04,Og04}. 
If the $L=2$ component consists of $l_3=0$ and $l_4=2$, the directions of $l_4$ and the spin 
for the single-particle state are also opposite, and this state becomes $0d_{3/2}$. 
This component makes $[(1s_{1/2})(0d_{3/2})]_{10}$ as shown in Fig.~\ref{fig:coupling}.
If $l_3=1$ and $l_4=3$, $[(0p_{3/2})(0f_{5/2})]_{10}$ is obtained in a similar way.
Higher configurations, having a set of larger orbital angular momenta of $l_3$ and $l_4$ 
possess small amplitudes, due to their large kinetic energies.
For this reason, the strong mixing of three $2p2h$ configurations in $^4$He 
can be understood from the selectivity of the operator of the tensor force.

\begin{figure}[t]
\centering
\includegraphics[width=13.2cm,clip]{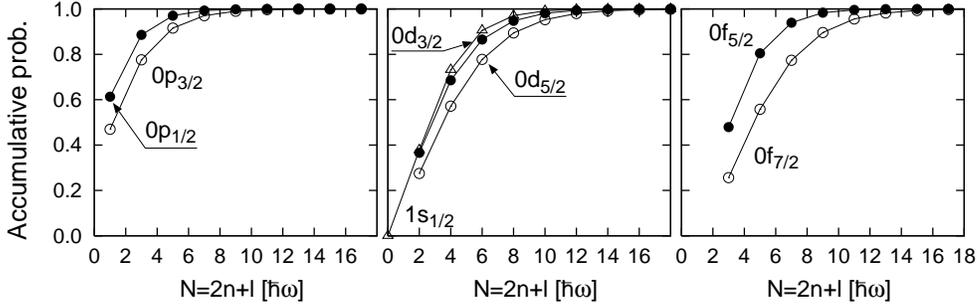}
\caption{Accumulative probabilities of the single-particle states for $^4$He in terms of HOWF for the $l_{\rm max}$=4 case.}
\label{fig:ho_expand}
\end{figure}

Here we discuss the relation of the spatial shrinkage of the single-particle states and the higher shell effect.
For this purpose, we expand the shrunk single-particle wave function in terms of HOWF in the same length parameter of the $0s$ state.
We calculate accumulative probability defined as the summation of the expansion probabilities up to the quanta $N=2n+l$.
The results are shown in Fig.~\ref{fig:ho_expand} for the $l_{\rm max}=4$ case.
A higher shell effect is seen, as $2p2h$ configurations require at least $14\hbar\omega$ excitations for the shrunk states.

\begin{figure}[t]
\centering
\includegraphics[width=9.0cm,clip]{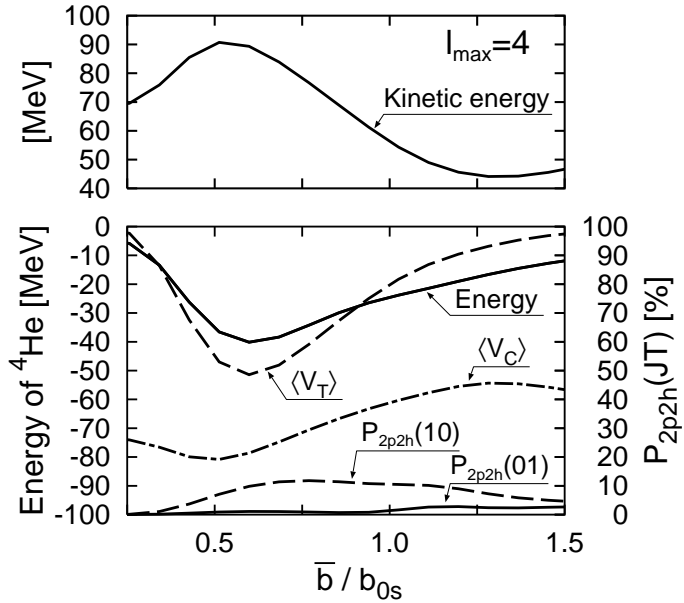}
\caption{The properties of $^4$He in the $l_{\rm max}$=4 case as a function of $\bar{b}$/$b_{0s}$. 
$P_{2p2h}(JT)$ are the probabilities of $2p2h$ components 
with $JT$ being the spin-isospin pair of the excited two nucleons.}
\label{fig:ene2}
\end{figure}

In Fig.~\ref{fig:ene2}, we plot the length parameter dependence of the solutions of $^4$He 
in order to see the effect of the spatial shrinkage more clearly, 
where length parameters, except for $b_{0s}$, are changed as a common value $\bar{b}$.
We plot the solutions as a function of the length parameter ratio $\bar{b}/b_{0s}$ in the $l_{\rm max}=4$ case, 
where $b_{0s}$ is fixed to 1.16 fm. 
The energy minimum is found to be near $\bar{b}=0.6\ b_{0s}$,
where $\langle V_T \rangle$ gives the largest contribution and the kinetic energy takes its maximum value.
In this case, $\langle V_T \rangle$ at the energy minimum is more than twice as large as 
in the ordinary shell model case with $\bar{b}/b_{0s}$=1.
For the $2p2h$ configurations, the $(J,T)=(1,0)$ pair is strongly mixed around the energy minimum. 
From these results, the tensor force is incorporated with a small length parameter for the excited nucleon states.

As shown in Table~\ref{tab:1}, the obtained wave functions give matter radii that are smaller 
than the experimental value (1.48 fm) by 0.2 fm. 
This is due to the fact that in addition to the overbinding of the system,
the length parameters of all single-particle states are smaller
than in the case of the single configuration of $(0s)^4$ for $^4$He with $b_{0s}$=1.4 fm, 
which reproduces the observed matter radii without the tensor correlation.

\section{Detailed analysis of the radial wave functions}
\label{sec:gauss}

In this section, we improve the radial wave functions of the particle states with the Gaussian expansion technique 
in order to obtain a quantitative description of the tensor correlation. 
We adopt the Gaussian expansion method\cite{Hi03,Ao95} for every excited particle state.
This technique has been shown to possess wide applicability for describing short-range correlations in 
few-body calculations and weakly bound states in neutron-rich nuclei.

\begin{figure}[t]
\begin{minipage}[l]{6.6cm}
{\makeatletter\def\@captype{table} 
\centering
\caption{The set of length parameters for the Gaussian basis functions.} 
\renc{\baselinestretch}{1.20}
\label{tab:4}
\centering
\begin{tabular}{l|l}\hline\hline
$N_\alpha$         &  set of $b_{\alpha,m}$~~($m=1,\cdots,N_\alpha$) \\ \hline
1                  &  0.7 \\
2                  &  0.6\,~0.8 \\
3                  &  0.6\,~0.8\,~1.0 \\
4                  &  0.6\,~0.8\,~1.0\,~1.4 \\
5                  &  0.6\,~0.8\,~1.0\,~1.4\,~2.0 \\ 
6                  &  0.4\,~0.6\,~0.8\,~1.0\,~1.4\,~2.0 \\ 
7                  &  0.4\,~0.6\,~0.8\,~1.0\,~1.4\,~2.0\,~2.5 \\ \hline
\end{tabular}
\makeatother}
\end{minipage}
\hspace*{0.1cm}
\begin{minipage}[r]{6.6cm}
\hspace*{-0.2cm}
\includegraphics[width=7.4cm,clip]{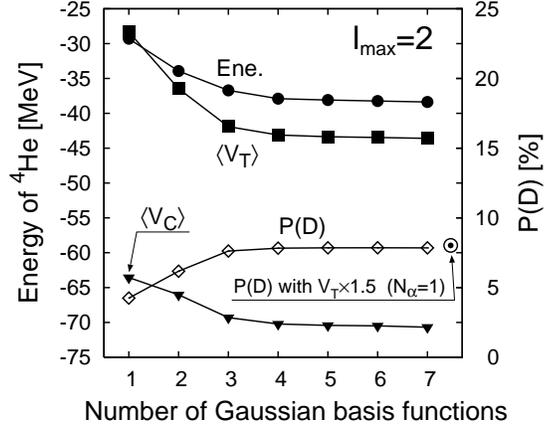}
\caption{The properties of \nuc{4}{He} obtained with the Gaussian expansion method.
The circle indicates $P(D)$ with a single Gaussian basis function realized by 
increasing the matrix elements of the tensor force by 50\%.}
\label{fig:Gauss}
\end{minipage}
\end{figure}

We now  present the results in the $l_{\rm max}$=2 case. 
Here, we fix $b_{0s}$ to 1.4 fm to obtain the appropriate matter radius of $^4$He in the present wave function,
as the results for the radius listed in Table~\ref{tab:2} are too small. 
We use the same number of basis functions, $N_\alpha$, for every $\alpha$ of the particle states.
The set of length parameters for the Gaussian basis functions in Eq.~(\ref{eq:Gauss2}) are listed in Table \ref{tab:4}.

From Fig.~\ref{fig:Gauss}, we confirm the convergence of the solutions with respect to $N_\alpha$.
In particular, the physical quantities almost converge within three or four bases,
where the bases with length parameters smaller than $b_{0s}$ are favored. 
It is also found that the Gaussian expansion gives a larger effect on $\langle V_T \rangle$ than $\langle V_C \rangle$.
In particular, $P(D)$ attains a value nearly twice as large as that in the single Gaussian case ($N_\alpha=1$).
The kinetic energy also increases about 20\% in the Gaussian expansion in comparison with the $N_\alpha=1$ case.
The matter radius converges to 1.39 fm.
We thus find that the Gaussian expansion succeeds in describing the radial wave functions for the excited particle states
and increases the tensor correlation for $^4$He.
In order to estimate this effect, we calculate $^4$He with a single Gaussian basis function. 
Doing so, we find that if we enhance the matrix elements of the tensor force by 50\%, the result of the Gaussian expansion is simulated.
This result implies that the reason for using the above enhanced tensor matrix elements in previous study\cite{My05} 
is confirmed from the improvement of the spatial description of the particle states.

\section{Properties of $^4$He with the tensor component of the nucleon-nucleon force}\label{sec:tens}

In this section we present the results obtained with TOSM for $^4$He using the tensor force
of the nucleon-nucleon interaction together with the central interaction modified by
accounting for the short-range repulsion.  
For this purpose, we first compare the matrix elements of the tensor forces obtained in various methods.

\subsection{The matrix elements of various tensor forces}

We compare the radial matrix elements of the tensor force $V_T=f_T(r) \cdot S_{12}$
for several interactions: the bare nucleon-nucleon interaction, AV8$^\prime$,\cite{Pu97} 
the Bonn potential with $\pi+\rho$,\cite{Ma87} and the $G$-matrix tensor interaction, AK, which is based on AV8$^\prime$.
In Fig.~\ref{fig:TF}, we show the radial dependence $f_T(r)$ of each tensor force for the triplet-even channel. 
We see that the short-range behavior depends very strongly on the choice of the nucleon-nucleon interaction. 
Contrastingly, the intermediate- and long-range behavior depends only weakly on the choice of the potential.

\begin{figure}[t]
\centering
\includegraphics[width=7.2cm,clip]{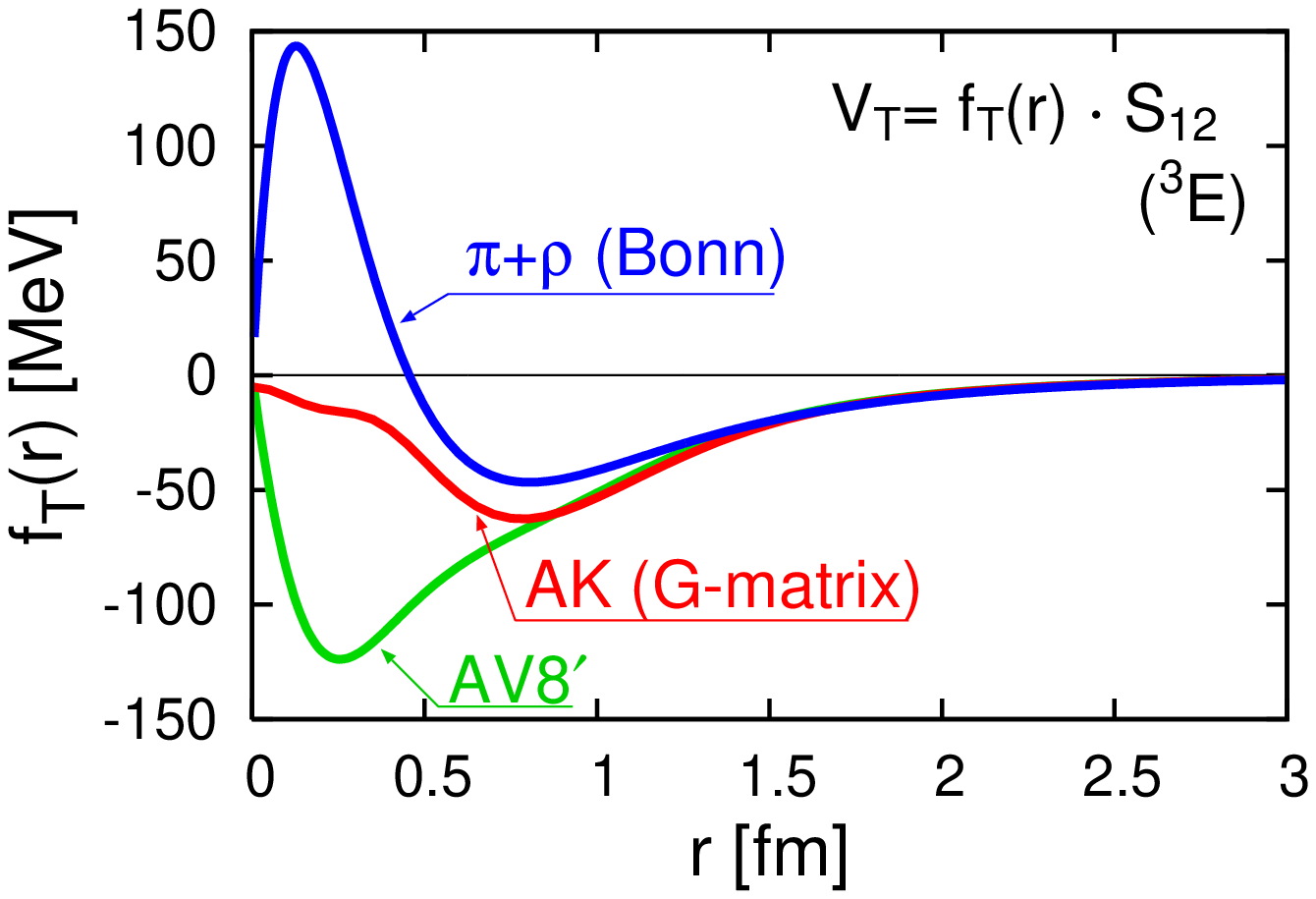}
\caption{Comparison of the tensor forces for the triplet-even channel.}
\label{fig:TF}
\end{figure}

\begin{figure}[t]
\centering
\includegraphics[width=6.7cm,clip]{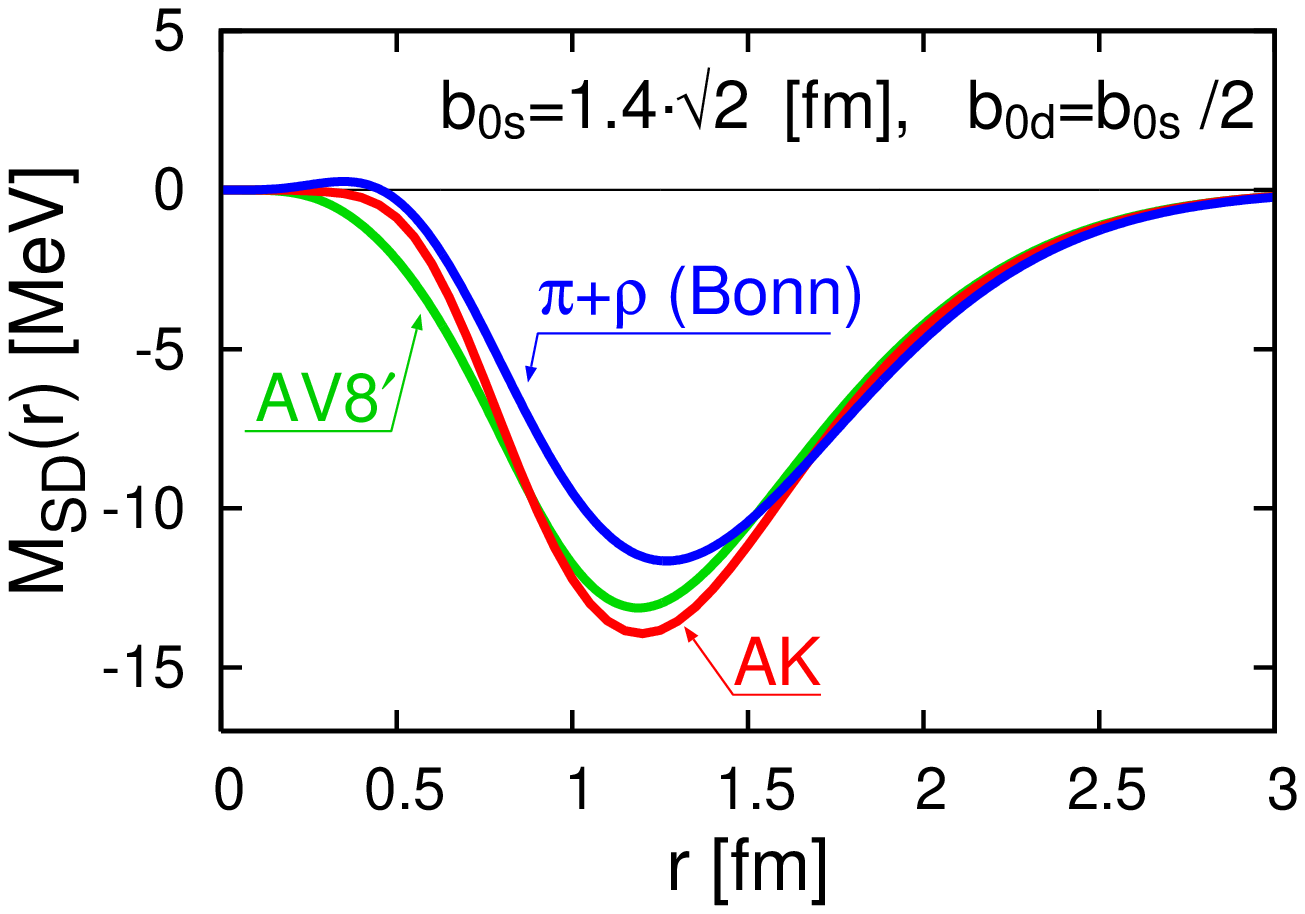}
\includegraphics[width=6.7cm,clip]{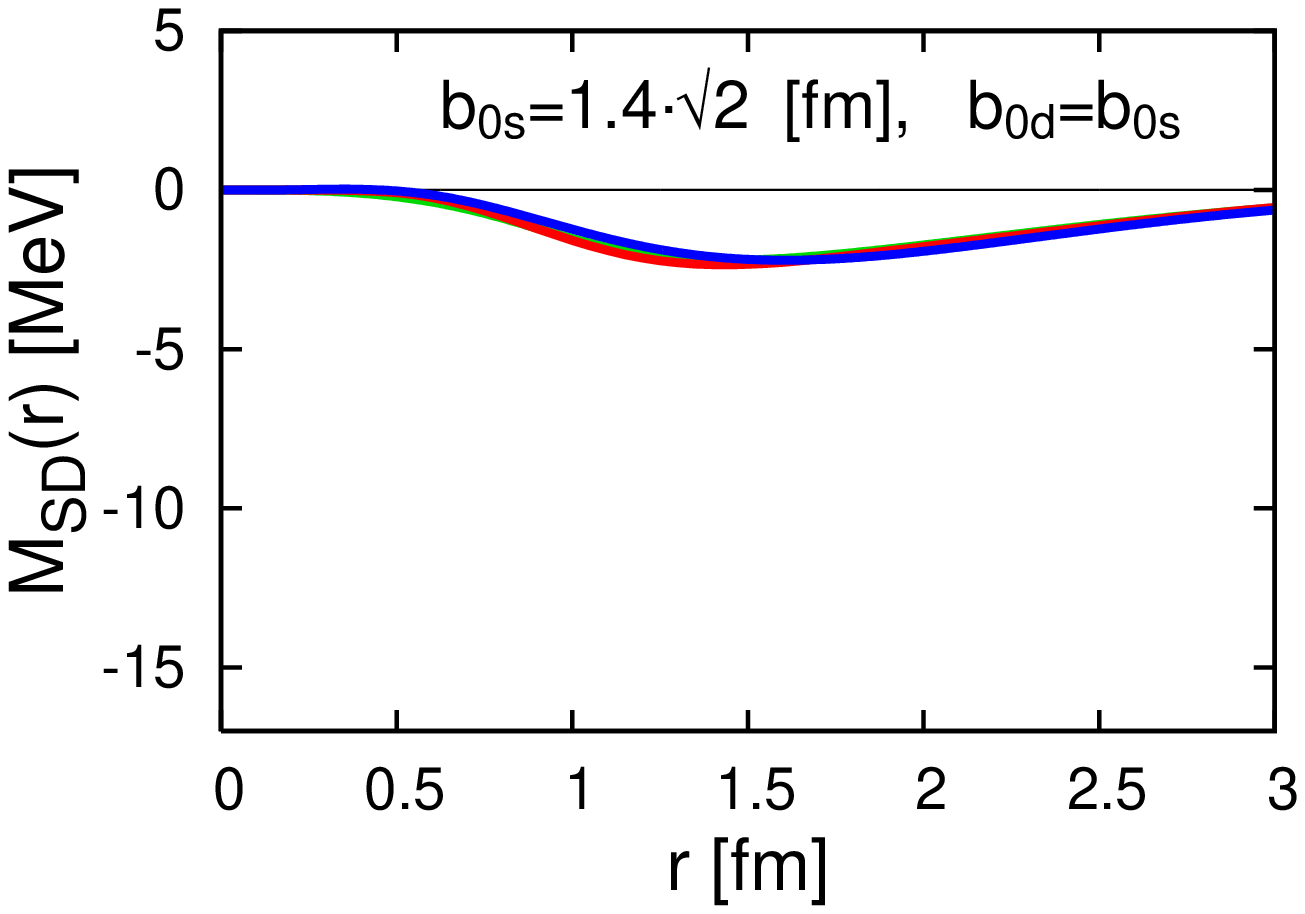}
\caption{Comparison of the integrands of the radial matrix elements of the tensor force in the cases of two values of $b_{0d}$.}
\label{fig:TF}
\end{figure}

We now compare the radial integrals for three kinds of tensor forces for the $S$-$D$ coupling in the relative motion 
defined by the following equations: 
\begin{eqnarray}
	I_{\rm SD}
&=&	\int_0^\infty dr \ M_{\rm SD}(r),
	\quad
	M_{\rm SD}(r) 
~=~	r^2 \phi_{0s}(r,b_{0s})\cdot f_T(r)\cdot \phi_{0d}(r,b_{0d}).
\end{eqnarray}
This matrix element is essential to explicitly describe the tensor correlation.
Here, $\phi_{0s}$ and $\phi_{0d}$ are the radial components of the Gaussian wave functions for the $s$ and $d$ waves
with the length parameters $b_{0s}$ and $b_{0d}$, respectively.
We choose $b_{0s} = 1.4 \cdot \sqrt{2}$ fm for the $0s$ wave function, 
where the factor $\sqrt{2}$ comes from the relative coordinate transformation of HOWF.
We consider two length parameters, $b_{0d} = b_{0s}/2$ and $b_{0d} = b_{0s}$, for the $0d$ wave function.  
The smaller length parameter $b_{0d}$, corresponds to the case of the largest matrix element of the tensor force,
 as discussed in the previous sections and the previous study,\cite{My05} and the larger length parameter corresponds to the case of the tensor force in the standard shell model.

In the left panel of Fig. \ref{fig:TF}, we depict the shorter range case for the radial integrands 
$M_{\rm SD}(r)$ in the three cases, which are seen to be very similar.  
The intermediate and long-range behavior is important for the matrix elements of the tensor force. 
The large difference in the short-range part is washed out due to
the $\Delta L=2$ transition of the tensor force for the relative orbital angular momentum. 
Due to the centrifugal barrier of the $D$ state, 
the contribution from short-range part becomes small in the tensor-force matrix elements.
This means that the coupling between the short-range correlation and the tensor correlation
is weak, and thus we can describe the tensor correlation explicitly in the shell model type method 
while we can renormalize the short-range correlation into the central force independently from the tensor force.  
In the right panel of Fig. \ref{fig:TF}, we also depict the longer range case.  Here, again the radial integrands are very similar.  
Further, we list the matrix elements for each tensor force, $I_{\rm SD}$, in Table \ref{tab:0}. 
We find that the three matrix elements are very similar, with differences on the order of 10\%.
We also see that the tensor matrix element for the shorter length parameter is much larger than 
that for for the larger length parameter, which corresponds to the case of the standard shell model.

\begin{table}[t]
\begin{center}
\caption{Radial matrix elements of the tensor force $I_{\rm SD}$ for the triplet-even channel in units of MeV.}
\renc{\baselinestretch}{1.20}
\label{tab:0}
\begin{tabular}{l|ccc}\hline\hline
                  &  AV8$^\prime$     &    AK        & $\pi+\rho$ (Bonn) \\
\hline
$b_{0d}=b_{0s}/2$ & $-$14.98   &  $-$15.04    & $-$13.30\\
$b_{0d}=b_{0s}  $ & ~\,$-$3.83 & ~\,$-$3.85   & ~\,$-$3.82\\
\hline
\end{tabular}
\end{center}
\end{table}

\subsection{The properties of $^4$He in TOSM}

We have studied the full treatment of the tensor force in the shell model type method by employing a sufficiently large number of angular momenta and radial wave functions using the $V$-type coordinates.  We have used the effective nucleon-nucleon interaction, GA.  We found good convergence in the case of intermediate angular momenta with considering the several values of the range parameters for the tensor force.  Because this is the first time that the tensor force has been treated fully,  while the short-range repulsion has to be treated somehow at present, the binding energy of $^4$He that we obtained is too large, and also, the size of $^4$He is too small. Here, we would like to adjust the parameters of the central interaction so that realistic values for the properties of $^4$He are obtained.  In the previous subsection, in addition, we showed that we are able to treat the bare tensor force obtained from nucleon-nucleon scattering and the matrix elements of tensor forces are found to be very similar, due to the large centrifugal potential.  Considering these points, we employ the bare tensor force for the study of $^4$He and use a phenomenological central interaction in TOSM.  

We adopt the tensor force and the LS force of AV8$^\prime$.  
For the central interaction, we begin with the central force of the GPT interaction, 
which is obtained by fitting the observed two-nucleon properties.
We then adjust the central part of GPT to improve the large binding energy, as explained in Table~\ref{tab:2}.
This is carried out by reducing the strength of the second range (attractive part) of the GPT's central force by 13\%
in order to fit the observed binding energy (28.3 MeV) for $^4$He.

For the $^4$He wave function, we choose $l_{\rm max}$=5 for high angular momentum states 
and use four Gaussian basis functions for every particle state, whose length parameters are 0.6, 0.8, 1.0 and 1.4 fm, respectively. 
These choices are sufficient in the present calculation.
We variationally determine the length parameter of the $0s$ state to be 1.42 fm.

\begin{table}[t]
\caption{The properties of \nuc{4}{He} in the $l_{\rm max}$=5 case with the use of the bare tensor force.
We list the contributions from each term in the Hamiltonian (in units of MeV) and the matter radius ($R_m$, in fm) and the probabilities (in \%) of each configuration labeling with the excited particle states.}
\renc{\baselinestretch}{1.15}
\label{tab:6}
\parbox{\halftext}{
\centering
\begin{tabular}{l|c}
\hline\hline
$E$                         & {$-$27.89}  \\ \hline
$\bra T\ket$                & {~~\,75.85}    \\
$\bra V_{C}\ket$            & {$-$44.57}   \\
$\bra V_{T}\ket$            & {$-$58.97}    \\
$\bra V_{LS}\ket$           & {~\,$-$0.98}      \\
$\bra V_{Clmb}\ket$         & { ~~\,\,0.78}        \\ \hline
$\bra H_{G}\ket/(1.5\hbar\omega)$& { \,1.011}  \\ \hline
$R_m$                       &  1.47           \\ \hline 
\end{tabular}
}
\parbox{\halftext}{
\centering
\begin{tabular}{c|c}\hline\hline
$P(D)$                      &\,\,~9.13  \\ \hline 
$0p0h$                      &\,85.02  \\
$(0p_{1/2})^2_{10}$         &\,\,~3.40  \\ 
$(1s_{1/2})(0d_{3/2})_{10}$ &\,\,~2.35  \\ 
$(0p_{3/2})(0f_{5/2})_{10}$ &\,\,~2.21  \\ 
$(0p_{1/2})(0p_{3/2})_{10}$ &\,\,~1.61  \\ 
$(0d_{5/2})(0g_{7/2})_{10}$ &\,\,~0.83  \\ 
$(0d_{3/2})^2_{10}$         &\,\,~0.60  \\ 
$(0p_{3/2})^2_{10}$         &\,\,~0.53  \\ \hline
\end{tabular}
}
\end{table}

The results for $^4$He are listed in Table~\ref{tab:6}. 
For the mixing probabilities, we sum the probabilities belonging to the same configurations having the same spin and isospin pair
with different radial components obtained in the Gaussian expansion.
It is found that $\langle V_T \rangle$ is approximately $-$60 MeV and $P(D)$ reaches approximately 9\%.
These results imply that the obtained wave function satisfactorily represents the characteristics of the tensor correlations.
The excitation of the spurious cm motion is weak, and its size is estimated to be approximately 0.4 MeV. 
The three $2p2h$ configurations are strongly mixed, and the $(0p_{1/2})^2_{10}$ component is most favored.

\section{Summary}\label{sec:summary}

We have studied the tensor correlation of $^4$He using a shell model type method with a $V$-type coordinate.  We have developed the tensor optimized shell model (TOSM) to fully treat the tensor correlation.  We have applied the TOSM to $^4$He, where we have employed the $(0s)^4$ basis and included up to the two-particle two-hole ($2p2h$) excitations with particle states up to intermediate angular momenta.  
Particle states are expressed in terms of Gaussian wave functions for each partial wave, 
while the length parameters are allowed to vary so as to minimize the total energy.  
For the study of the tensor correlation, we first considered the $G$-matrix interaction based on the AV8$^\prime$ interaction\cite{Ak04,Ik04}
and replace the central part of the interaction by those of the GPT\cite{Go70}.

We first calculated $^4$He by including the intermediate high angular momentum states up to $l_{\rm max}$ successively, while employing only one Gaussian basis function for each orbit.
We found that the spatial size of the particle states are significantly shrunk in order to optimize the tensor correlation for $^4$He.  
The Gaussian length parameters of high partial waves are found to be about 60 \% of that of the $0s$ state.
The energy gain from the tensor force greatly as a function of $l_{\rm max}$
and exhibits convergence near $l_{\rm max}=4$.  
We found that three particular $2p2h$ states admix to the $(0s)^4$ component due to the characteristic features of the tensor force.  
We then expressed the radial wave function in terms of the superposition of Gaussian basis functions with several length parameters. 
We found that approximately three Gaussian basis functions are sufficient to express the radial wave functions, 
and the energy gain resulting from the use of various Gaussian wave functions is about 
50\% in comparison with the case of one Gaussian wave function.  
Hence, we have succeeded in developing a shell model type model based on the $V$-type coordinate, TOSM, 
which is capable of treating a strong tensor force.

We then studied $^4$He with TOSM using the bare tensor force AV8$^\prime$ and the short-range modified central interaction GPT. 
We were able to determine the ground state properties of $^4$He with a modification of the intermediate range of GPT central interaction 
and thereby reproduced the binding energy.
The expectation value of the tensor force is very large, about $-60$ MeV, and the $D$-state probability is approximately 10\%.

We developed the TOSM, with which we are able to treat the bare tensor force, 
with the hope of investigating the tensor correlations in heavy nuclei.  As the next step we should develop a method 
to incorporate the short-range repulsion so that we are able to treat intermediate and heavy nuclei using the nucleon-nucleon interaction.
We plan also to study the effects of the tensor correlation on the structures of the nuclei in the neighborhood of $^4$He.  
Particularly interesting are the scattering phenomena of $^4$He+neutron system ($A=5$) and 
the structures of $A=6$ systems, $^6$He, $^6$Li and $^6$Be.

\section*{Acknowledgements}

The authors would like to thank Prof. H. Horiuchi and Prof. T. Yamada for fruitful discussions and encouragement.  
This work was supported by a Grant-in-Aid from the Japan Society for the Promotion of Science (JSPS, No. 18-8665) and the 21st Century COE program, Center for Diversity and Universality in Physics at Kyoto University. 
This work was performed as a part of the Research Project for Study of Unstable Nuclei from Nuclear Cluster Aspects (SUNNCA) at RIKEN.
Numerical calculations were performed on the computer system at RCNP.

\end{document}